# Integrated lithography to produce complex structures for spectral engineering

**Mária Csete,\* Áron Sipos, Anikó Szalai and Gábor Szabó**
Department of Optics and Quantum Electronics, University of Szeged, H-6720, Szeged, Dóm tér 9, Hungary

E-mail: mcsete@physx.u-szeged.hu

**Abstract.** An integrated interference and colloid sphere lithography (IICL) is presented to produce complex plasmonic structures consisting of wavelength-scaled periodic arrays of nano-objects with arbitrary array symmetry and controllable nano-scaled sub-structure. The IICL method is based on illumination of colloid sphere monolayers by interference patterns synchronized with sphere arrays along arbitrary crystallographic directions. This nano-kaleidoscope method enables to tune four structure parameters independently: the symmetry and characteristic periodicity of the interference pattern might be varied by the wavelength, number and angle of incidence of the interfering beams; the colloid-spheres' diameter-scaled distance between the nano-objects is controllable by the relative orientation of the interference pattern with respect to the hexagonal lattice of colloid spheres; the size of individual nano-objects is determined by the colloid-spheres diameter and by the light wavelength and is influenced by power density; the sub-structure size-parameter sensitively depends on the polarization state and can be tuned with the nano-object size simultaneously. Finite element method is applied to demonstrate the capabilities of IICL, and the impact of the resulted complex plasmonic patterns on spectral properties of thin gold films. The possibility to realize spectral engineering with predesigned rectangular arrays of hole-doublets that may be produced uniquely by IICL is shown.

## 1. Introduction

In last decades significant progress has been made in miniaturization of optoelectronic devices due to advancements in material science. The optical wavelength-scaled periodic structures, e.g. photonic crystals [1], and nano-scaled metal patterns, e.g. nano-plasmonic materials [2], are at the heart of these devices. Periodic arrays of nano-objects may offer novel solutions for many scientific and technological problems. The plasmonic spectral engineering concept is based on the sensitivity of complex noble metal structures' spectra to the geometry of individual nano-objects and their arrays. Designing controlled complex plasmonic structures is likely to provide revolutionary methods for application areas such as bio-sensing [3].

High resolution features in arbitrary pattern can be generated by classical scanning beam lithography methods, but application of these techniques is limited caused by the related high costs and slow processing rate [4]. In practical applications there is a great demand for low cost and rapid technologies.

Various laser-based material treatment methods with widely tunable experimental parameters are available in science and in industry. These methods are parallel processes, i.e. make it possible to pattern large-scale areas at once. Periodic structures may be fabricated on macroscopic substrates by high fluence laser pulses, applying mirrors to realize two-beam interference [5]. Our previous studies have shown that sub-micrometer periodic metal stripes of nano-grains can be produced this way [6]. Caused by the inherent diffraction limit in all laser-based techniques, generation of nano-scaled structures is realizable only via hardly controllable self-organized processes [7].

Colloid sphere lithography is a suitable technique to produce sub-wavelength objects, and is widely applied to fabricate periodic nano-particle arrays with precisely controlled shape, size, and inter-particle spacing on large areas [8].

The laser-based colloid sphere lithography is capable of overcoming the diffraction limit [9, 10]. The strong near-field enhancement achievable, when dielectric colloid spheres are illuminated with coherent light, is determined by the spheres' diameter and by the wavelength, while the plane of maximal intensity depends also on the relative index of refraction [11]. In case of perpendicular incidence the intensity distribution is determined by optical resonances on spheres larger than the wavelength and by near-field effects under spheres with sub-wavelength diameter [12].

It is possible to fabricate controllable sub-structures inside the treated areas with versatile symmetry applying oblique incident beams, when the diameter-wavelength ratio is appropriate [13, 14]. It was also demonstrated that silica colloid sphere masks can be used to produce hole-arrays as a first step, and then the templates made of these hole-patterns make it possible to produce metal nano-dot arrays via lift-off process [15].

The effect of metal spheres has been less investigated until now, but is promising due to the enormous field enhancement accompanying the localized surface plasmon resonance, which might be tuned by the diameter and by the dielectric properties of media around the spheres [16]. In presence of substrates the largest enhancement was observed, when metal colloid spheres were arrayed on metal surfaces [17, 18]. The near-field confinement under metal spheres is more strongly dependent on the polarization and angle of incidence than on dielectric spheres, which makes high-resolution surface plasmon assisted nano-patterning possible [19].

The advantage of colloid sphere based photonic methods is the very high uniformity of the sub-wavelength objects, while the commensurability and fixed ratio of the *a* object size and the *d* inter-object distance inherently limits the degrees of freedom available in geometrical parameters' variation. The hexagonal arrangement in closely packed sphere monolayers defines the symmetry properties of the structures that can be produced and therefore limits the geometry-dependent spectral properties, however engineering of these properties is crucial in different application areas. There are tremendous efforts in recent colloid sphere lithography to overcome these barriers, among them are modification of the spheres' assembly process, and application of templates to predefine their deposition [20]. These techniques are multi-step chemical procedures, which require special experimental infrastructure. Development of dominantly optical methods that necessitate the lowest number of consecutive steps and extensively available apparatus is at the core of recent applied colloid sphere based nanotechnology.

We propose a novel Integrated Interference and Colloid sphere Lithography (IICL) method having a potential to generate complex e.g. plasmonic structures consisting of wavelength-scaled arrays of arbitrary uniform nano-features, as holes or particles, on large areas at low-costs. Our further purpose is to show that the novel lithography method is capable of tuning the dominant micro- and nanoscopic parameters independently, which determine the spectral properties of complex plasmonic structures. The final purpose of the present paper is to show an example of spectral engineering, which is realized via appropriately designed complex plasmonic structures.

## 2. Integrated interference and colloid sphere lithography: method and capabilities

In IICL, intensity distribution with appropriate symmetry and periodicity is applied to select arbitrary arrays from colloid spheres in hexagonal monolayers. Illumination by an interference pattern completely substitutes the chemical pre-treatment procedures or application of templates that might result in non-hexagonal colloid sphere arrays via multi-step processes. Implementation of the IICL method requires the synchronization of an interference pattern with a colloid sphere monolayer, and perfectness either of the intensity distribution and the monolayer. An optimized relative positioning allows to maintain periodic intensity modulation and near-field enhancement simultaneously, thus making possible the near-field concentration of the periodic intensity distribution (Figure 1).

IICL is a parallel technique, as macroscopic surface parts might be patterned at once, where the periodic intensity distribution matches colloid-sphere arrays in a perfect monolayer. The method combines the advantages of both techniques: the symmetry and periodicity of the wavelength-scaled interference modulation defines a periodic pattern, while the intensity concentration in the near-field of colloid spheres occurs only within areas considerably smaller than the wavelength.

In addition to the available resolution an important figure of merit in lithography is the available degree of freedom, which specifies the number of independently tunable geometrical parameters. First we will present, how the geometrical size parameters can be controlled in IICL, and show how this novel integrated lithography method combines all capabilities of the interference and colloid sphere lithography, resulting uniquely in four degrees of freedom.

In interference lithography the number and incidence angle of the interfering beams determine the symmetry properties of the fabricated patterns. Two-beam interference results in linear arrays, while multiple-beam interference is capable of producing complex patterns with arbitrary symmetry. The *first* characteristic geometrical size parameter in IICL is the mesoscopic $p$ periodicity (100 nm – 1000 nm) of the nano-object array tunable by the $\lambda$ wavelength and $\theta$ angle of incidence, in accordance with interference lithography methods. The extension of the area, where nano-objects might be fabricated inside interference patterns is synchronously controllable by the wavelength and angle of incidence, and is influenced by the power density. It is possible to define and control a secondary pattern by tuning these parameters in experiments. In IICL the interference pattern is wavelength scaled, as in interference lithography, but the resulted pattern has a fine structure due to the concentration of EM-field by colloid spheres.

The *second*, nano-scaled geometrical size-parameter is the $t$ inter-object distance (10 nm – 1000 nm), which is commensurate with the colloid sphere's $d$ diameter. This degree of freedom appears as a result of interference and colloid lithography integration. It depends on the relative orientation of the interference pattern with respect to the colloid sphere monolayer. As the relative orientation might be varied semi-continuously, there are countable infinite number of values that $t$ might take on (Figure 1). Under the same experimental condition, i.e. applying light with the same wavelength and colloid spheres having the same material and diameter, various arrays might be fabricated simply by varying the illumination directions.

The relative orientation of the interference pattern with respect to the monolayer, and the angle of incidence has to be tuned together. In this paper two significantly different geometries are introduced for the case of two-beam interference, which have the smallest possible $t$ inter-object distance. These two representative examples are nominated as geometry-I and geometry-II, and the corresponding quantities are indicated by roman numbers I and II in Figure 1.

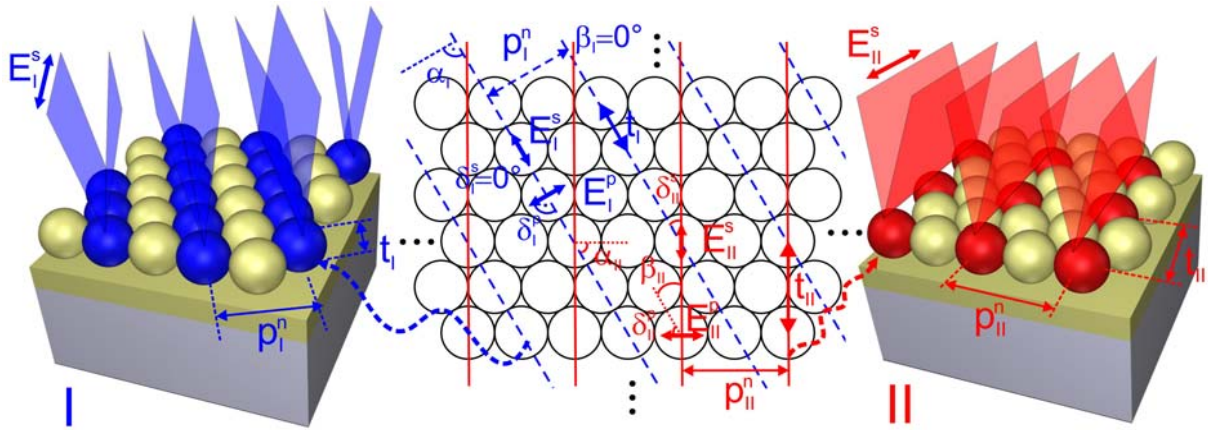

Figure. 1. Schematic drawing about the "nano-kaleidoscope" method presenting two possible relative orientations in IICL. The index $i$=I refers to geometry-I; while $i$=II refers to geometry-II. The relevant size parameters are the $p_i$ period and $t_i$ inter-particle distance; integer values of $n$ refer to scaling of periodicity as $p_I^{(n)} = n \cdot \sqrt{3}/2 \cdot d$ and $p_{II}^{(n)} = n \cdot d/2$. The angles $\alpha_i$, $\beta_i$=90°−$\alpha_i$ and $\delta_i^{s/p}$ refer to the relative orientations of the plane of incidence, interference maxima, and $E_i^{s/p}$-field oscillation in s- and p-polarized light with respect to the (1, 0, 0) crystallographic direction, respectively. The spheres under interference maxima are highlighted in blue (I) and red (II) colors on the insets.

All directions are measured with respect to the array of closely packed colloid spheres, which is the (1, 0, 0) direction in two-dimensional hexagonal lattices of colloid spheres. When the angle between the (1, 0, 0) direction and the plane of incidence is $\alpha_i$ the interference pattern is oriented under $\beta_i$= 90° − $\alpha_i$ in case of two-beam interference.

The direction of **E**-field oscillation is specified by $\delta_i^{s/p}$ angle, i.e. s- and p-polarized light illumination is qualified by $\delta_I^s = 0°$ and $\delta_I^p = 90°$ in geometry-I (Figure 1, highlighted in blue), and $\delta_{II}^s = 30°$ and $\delta_{II}^p = 60°$ in geometry-II (Figure 1, highlighted in red).

When the interference pattern is parallel to the array of closely packed colloid spheres, which is the (1, 0, 0) direction, i.e. $\beta_I = 0°$, the inter-object distance equals with the spheres diameter $t_I = d$. The inter-array distance, i.e. the periodicity is: $p_I^n = n \cdot \sqrt{3}/2 \cdot d$, which is larger than the distance of nano-objects in case of homogeneous illumination (HI), always when $n>2$. The angles of incidence appropriate to generate interference pattern with periodicities corresponding to different values of $n$ in geometry-I are:

$$\theta_I^n = \arcsin\left(\frac{\lambda}{n \cdot \sqrt{3} \cdot d}\right). \quad (1)$$

When the interference pattern is rotated with respect to the array of closely packed colloid spheres by $\beta_{II} = 30°$, the produced nano-features are more closely packed in between arrays. The inter-object distance is larger than the sphere diameter $t_{II} = \sqrt{3} \cdot d$, while the inter-array distance, i.e. the periodicity is scaled by the half-diameter as $p_{II}^n = n \cdot d/2$, which is larger than the distance between nano-objects in case of homogeneous illumination, always when $n>2$.

The angles of incidence appropriate to generate interference pattern with periodicities corresponding to different values of $n$ in geometry-II are:

$$\theta_{II}^n = \arcsin\left(\frac{\lambda}{n \cdot d}\right). \quad (2)$$

When $n$ is an even number, the periodicity is scaled by the distance of $\sqrt{3} \cdot d$ in geometry-I, and by the $d$ diameter in geometry-II, as shown in present study. The advantage of cases corresponding to even $n$ numbers is that the resulted nano-objects are aligned in linear arrays perpendicularly to the interference pattern, while odd $n$ numbers cause that the nano-objects are shifted with respect to each other alongside neighboring interference maxima. The only barrier in IICL is that the $\theta$ angle of incidence is limited by the shade effect of neighboring spheres, and the maximal value is 45°, when they are dielectric, while the incidence angle may reach even 90° in case of metal spheres.

Due to these capabilities, IICL is a "nano-kaleidoscope" method, as the diffraction on the produced complex patterns might be controlled by the $\alpha_i / \beta_i$ rotation angles. An important advantage of IICL based on metal spheres is that the inter-particle coupling might be optimized synchronously, since the distance between the particles illuminated by interference maxima is also tuned by $\theta$ angle of incidence, when the tilting is varied, as we show in Section 4.

The *third* geometrical size parameter is the nanoscopic $a$ diameter of the treated areas (1-10 nm, 100 nm), which is commensurate with the FWHM of the intensity distribution under colloid spheres. This parameter is determined by the spheres $d$ diameter and by the $\lambda$ wavelength and is influenced by the applied power density. The related degree of freedom originates from colloid sphere lithography. Forward step in IICL is that the nano-scaled features are arrayed according to the interference pattern.

In case of polarized light illumination a characteristic sub-structure appears, which might be qualified by a *fourth* $d_0$ sub-structure size-parameter. Caused by symmetry reasons this sub-structure is aligned alongside the interference pattern in case of s-polarized light, while p-polarized light illumination results in a sub-structure aligned perpendicularly to the interference pattern. The $d_0$ size parameter is tunable simultaneously with $a$, by varying the colloid sphere diameter, wavelength, and power density. Appearance of arrays composed of these sub-structures is a synergic effect, as their properties are predetermined by the complete experimental parameter setting, i.e. the corresponding degree of freedom is not completely independent. Although the type of polarization has analogous effect in standard colloid sphere lithography, IICL uniquely provides possibility of choosing different relative orientations between the sub-structure and interference pattern.

In laser-based colloid sphere lithography the illumination results in exploitation of the spheres, while in IICL the spheres illuminated by beams with power density below certain threshold have to be removed in a single post-treatment step. The selective removal might be promoted by an appropriate chemical layer, but such interfacial layer is not required in all colloid sphere and substrate systems. IICL method can be used directly to produce concave objects, e.g. holes in different substrates or thin films, and these concave objects might be converted into nano-particles via standard lift-off lithography. We have selected metallic film covered substrates for the presentation of IICL capabilities, although the method is applicable to fabricate any kind of targets.

The most significant spectral effects originating from plasmonic phenomena determined by the surface structure of metal films or metal substrates are well described in the literature. When the propagating EM-modes' wavelength is commensurate with the periodicity of the nano-object array, extrema appear on the absorption and transmission spectra of nano-particle- and hole-arrays [21, 22]. With IICL metal films might be directly structured by illuminating colloid sphere monolayers arrayed on the top of them, so applying this method it becomes possible to design complex patterns with extraordinary transmission at arbitrarily predefined wavelengths [23]. Moreover, due to the four degrees of freedom, IICL might be the most convenient method to generate complex arrays of nano-objects, i.e. pre-designed plasmonic nano-clusters that seem to be the most promising in several application areas [24].

### 3. Finite Element Method to demonstrate IICL

We used Finite Element Method (FEM) to determine the EM-field distribution under gold colloid spheres arrayed on 45 nm thick gold film coated NBK7 substrates.

The computations were performed by applying the Radio Frequency module of Comsol Multiphysics software package (COMSOL AB). Special 3D models were created, which make it possible to illuminate colloid sphere monolayers with infinite lateral extension by arbitrary number of beams with arbitrary wavelength, polarization state and angle of incidence. The normalized (time-averaged) electric-field was investigated to determine the near-field enhancement on the substrate surface under the colloid spheres, and to predict the expected *a* nano-object size based on the FWHM, and the $d_0$ sub-structure parameter based on the splitting of the lateral intensity distribution.

As a first step, the intensity distribution under colloid sphere monolayers was computed for the case of homogeneous illumination (Figure 2, 3/a, d). As a second step, the angles of incidence appropriate to result in interference patterns with desired periodicities were theoretically determined for specific cases of geometry-I and geometry-II in IICL (Figure 2, 3/b, c, e, f), and the interference pattern was synchronized with the colloid sphere monolayers. The characteristic parameters, such as near-field enhancement, FWHM, splitting in intensity distribution, determined for the case of homogeneous illumination of single spheres and monolayers, and applying different parameter settings in geometry-I and in geometry-II in IICL are compared in Figure 4.

Finally, the spectral effects of hole-arrays in thin gold films consisting of stand-alone holes, and hole-doublets in hexagonal pattern, which might be produced by HI, and in rectangular pattern, which can be exclusively fabricated by IICL, were analyzed. The purpose of this comparative spectral study was to prove that fine spectral tuning might be realized by hole-arrays produced via IICL due to geometry-dependent nano-plasmonic effects (Figure 5). The wavelength was tuned from 300 to 900 nm in these spectral studies, and the wavelength dependent optical parameters were taken into account by Cauchy formulas of the dielectrics, and by loading tabulated data-sets of the dielectric parameters for gold.

### 4. Results and discussions

Figures 2 and 3 indicate the effect of gold colloid sphere monolayers on the intensity distribution at the surface of 45 nm thick gold film coated substrates in case of illumination by s-polarized light. These computation results prove that the gold spheres can act as efficient nano-lenses capable of light concentration in the near-field, when they are arrayed in monolayers.

*4.1. Effect of homogeneous illumination*

Figures 2a, d and 3a, d present the reference cases, i.e. when the monolayers made of 250 nm and 100 nm diameter colloid spheres are illuminated by perpendicularly incident homogeneous light. As a result of illumination by linearly polarized light, split intensity distribution is observable along the *E*-field oscillation direction. According to this, a sub-structure manifesting itself in hole-doublets may develop under each of the colloid spheres.

The intensity distribution under the monolayer is influenced by $\delta_I$ angle between the (1, 0, 0) orientation and the *E*-field oscillation direction already in case of illumination by perpendicularly incident homogeneous beam (Figure 2a-to-d).

The *E*-field oscillates along the spheres closely packed inside arrays, according to $\delta_I^s = 0°$ in Figure 2a, while the *E*-field oscillation is along spheres arrayed at a distance of $t_{II} = \sqrt{3} \cdot d$ in Figure 2d and Figure 3a, d, according to $\delta_{II}^s = 30°$. The large-scale intensity distribution has hexagonal symmetry in both geometries, but the complex patterns consist of a hexagonal array of doublets possessing split along different *E*-field oscillation directions.

*4.2. Effect of illumination by interference patterns*
Light illumination by two interfering beams results in a periodic intensity modulation determined by the wavelength and incidence angle. As s-polarized light was applied in present study, the relative orientations of the *E*-field oscillation with respect to the (1, 0, 0) direction were $\delta_I^s = 0°$ in geometry-I (Figure 2a-c) and $\delta_{II}^s = 30°$ in geometry-II (Figure 2d-f, Figure 3).

First it was analyzed, how the $\beta_i$ relative orientation between the monolayer and the interference pattern influence the intensity distribution under 250 nm gold colloid spheres, when their monolayer is arrayed on 45 nm thick gold film covered NBK7 substrate. The monolayers were illuminated by two interfering 532 nm light beams in geometry-I and in geometry-II (Figure 2a-c to d-f). The tuning of the *p* parameter by the angle of incidence is also shown, when s-polarized light is incident at polar-angles according to *n* = 2 and *n* = 3 cases (Figure 2b-to-c); *n* = 4 and *n* = 6 cases (Figure 2e-to-f).

When we illuminate colloid spheres in geometry-I and geometry-II, the different arrangement of the spheres aligning at $t_I = d$ and $t_{II} = \sqrt{3} \cdot d$ distances under the interference maxima with finite width results in different complex patterns formation.

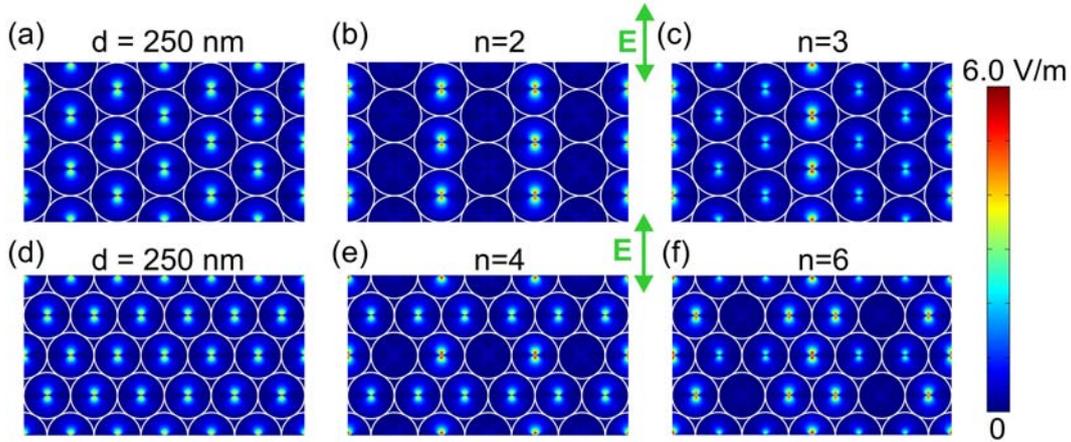

Figure 2. Near-field distribution under monolayers of 250 nm diameter gold spheres illuminated by s-polarized light having a wavelength of 532 nm (arrows indicate the direction of polarization), (a-c) in geometry-I and (d-f) in geometry-II. The intensity distribution of the incoming beam is (a, d) homogeneous; and modulated by an interference pattern corresponding to (b) *n* = 2: $\theta_I^2 = 37.90°$, (c) *n* = 3: $\theta_I^3 = 24.18°$, (e) *n* = 4: $\theta_{II}^4 = 32.14°$; and (f) *n* = 6: $\theta_{II}^6 = 20.77°$.

In geometry-I the intensity distribution under the spheres indicates doublets along closely packed sphere arrays, as both the interference maxima and the *E*-field oscillation direction are parallel to the (1, 0, 0) orientation *($\beta_I = \delta_I^s = 0°$)*. The comparison of the near-field pictures taken about 433 nm periodic pattern appearing at $\theta_I^2 = 37.90°$ incidence angle corresponding to *n* = 2 (Figure 2b), and about the 649.5 nm periodic pattern developing at $\theta_I^3 = 24.18°$ incidence angle corresponding to *n* = 3 (Figure 2c) indicates that smaller incidence angle results in linear nano-object array with larger *p* period. Moreover, in case of smaller incidence angle there is an EM-field enhancement along two neighboring arrays, caused by the larger extension of the interference maxima. As a result, a linear secondary pattern develops and as a synergic effect, a complex pattern of "nano-butterflies" with wings opened along the *E*-field oscillation direction appears along incidence angle dependent number of linear arrays consisting of closely packed spheres in geometry-I (Figure 2a-c).

When a monolayer of 250 nm gold spheres is illuminated by s-polarized light in geometry-II, significantly different complex pattern appears. The *p* periodicity might be tuned by the incidence angle, e.g. 500 nm periodic pattern develops at $\theta_{II}^4 = 32.14°$ corresponding to *n* = 4 (Figure 2e), and 750 nm periodic pattern appears at $\theta_{II}^6 = 20.77°$ corresponding to *n* = 6 (Figure 2f). The intensity modulation with finite extension results in a hexagonal secondary pattern, caused by the $\beta_{II} = 30°$ relative orientation of the interference pattern and $\delta_{II}^s = 30°$ *E*-field oscillation with respect to (1, 0, 0) direction.

By decreasing the angle of incidence, the extension of the secondary pattern increases simultaneously with the increase of the periodicity. The synergic effects result in vertically arrayed "nano-flowers" in geometry-II, but the secondary pattern is more pronounced at similar incidence angles than in case of geometry-I (Figure 2b-to-e and c-to-f). According to this, patterning might be performed over a more extended area and results in more complex nano-clusters in geometry-II.

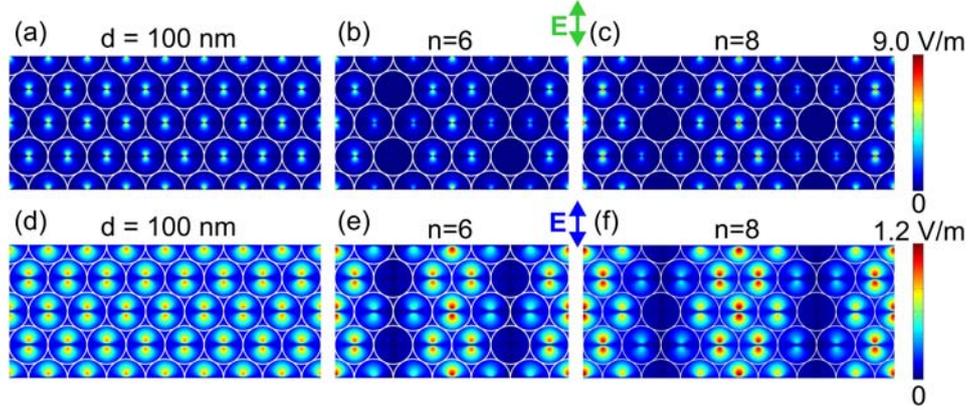

Figure 3. Near-field distribution under monolayers of 100 nm diameter gold spheres illuminated in geometry-II by light having a wavelength of (a-c) 532 nm, and (d-f) 400 nm (arrows indicate direction of polarization). The intensity distribution of the incoming beams is (a, d) homogeneous, and modulated by an interference pattern corresponding to $n = 6$ at (b) $\theta_{II}^{6}(532nm) = 62.46°$, and (e) $\theta_{II}^{6}(400nm) = 41.81°$ angles of incidence; and $n = 8$ at (c) $\theta_{II}^{8}(532nm) = 41.68°$, (f) $\theta_{II}^{8}(400nm) = 30.00°$ angles of incidence.

Figures 3a-f show the effect of 100 nm diameter gold colloid spheres on the intensity distribution. The role of the colloid spheres size was analyzed by illuminating 250 nm and 100 nm diameter spheres by interference patterns of two 532 nm beams, incident under $\theta$ angles corresponding to the same $n = 6$ case in geometry-II. The comparison of Figure 2f to Figure 3b shows that the $p$, $a$ and $d_0$ geometrical parameters might be tuned by the colloid spheres diameter. Smaller colloid spheres result in smaller size-parameters, and the secondary pattern develops on a relatively larger area at the same $n$ values (Figure 3 and 4).

The effect of illuminating light wavelength was investigated by shining 100 nm diameter colloid spheres by 532 nm and 400 nm beams. The comparison of the near-field pictures shows that the large-scale normalized *E*-field distribution is similar at different wavelengths under the colloid sphere monolayers (Figure 3a-c to 3d-f). Patterns possessing analogous periodicity appear, in case the beams are incident at $\theta$ angles corresponding to the same $n = 6$ and $n = 8$ cases in geometry-II (Figure 3b-to-e and c-to-f). Important quantitative difference is that the enhancement is significantly larger in case of illumination by 532 nm wavelength light, which is close to the resonance wavelength of gold.

*4.3. Quantitative near-field study*
A detailed quantitative near-field study was performed to inspect the effect of the IICL geometry, sphere diameter and wavelength (Figure 4a-d). The *E*-field enhancement is the largest, when a stand-alone gold sphere is illuminated, and the intensity is much lower, when a monolayer built up from touching gold colloid spheres is shined, according to the literature [17, 18]. Applying two interfering beams, the smaller incidence angles result in increased periodicity, which is accompanied by less efficient coupling between the gold particles. As a result, the enhancement is larger in IICL, than in case of homogeneous illumination of colloid sphere monolayers.

The FWHM of the intensity maxima is commensurable in all cases, i.e. when single sphere is illuminated and when monolayers are illuminated by performing HI and IICL. The degree of splitting in intensity distribution, i.e. the distance between the two split intensity maxima is also commensurate in case of HI and IICL, due to the parallelism of the *E*-field oscillation direction to the interference maxima during s-polarized light illumination.

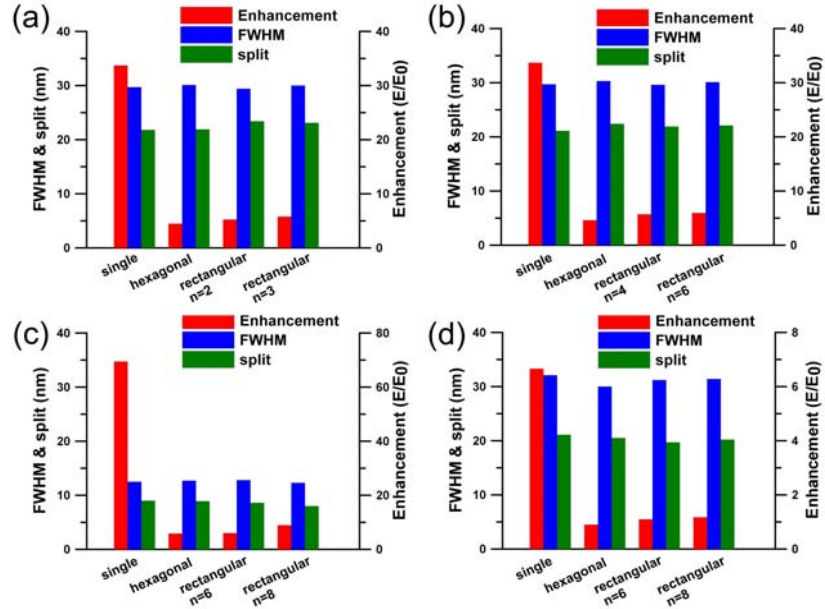

Figure 4. Enhancement, FWHM and splitting values, when $d = 250$ nm spheres are illuminated by 532 nm wavelength light beams (a) in geometry-I, and (b) in geometry-II; and when $d = 100$ nm spheres are illuminated in geometry-II by (c) 532 nm and (d) 400 nm wavelength light beams. Please notice that the scale in (d) inset on the right side is 10 times smaller according to much smaller enhancement.

*4.4. Application of complex plasmonic structures produced by homogeneous illumination and via integrated interference and colloid sphere lithography for spectral engineering*

The geometrical parameters of the holes developing as a result of HI and IICL are extrapolated based on the time-averaged *E*-field distribution at the colloid spheres' and gold film interface in present work. The material response, i.e. the heat-distribution inside the illuminated areas, and the resulted geometry of the produced holes are strongly influenced by the power density applied in experiments. The approximation of the hole-diameter by the FWHM, and the distance between individual holes by the splitting in intensity distribution is based on the theoretical proof of similar resistive heating distribution and on the experimental evidence of structure parameters' sensitivity to the power density. The spectral effect of nano-hole arrays developing as a result of 532 nm wavelength illumination of 100 nm diameter gold colloid spheres was analyzed. The studied unit cells are indicated by red rectangles on the schematic drawings about the colloid sphere monolayers (Figure 5a, b). In Figure 5a size parameters of $d = 100$ nm and $t = \sqrt{3} \cdot d = 173\,nm$ reveal to the diameter of colloid spheres in a hexagonal lattice (Figure 3a), while in Figure 5b size parameters of $p = 300$ nm, $t = \sqrt{3} \cdot d = 173\,nm$ are in accordance with periodicities in geometry-II (Figure 3b). The diameter of stand-alone holes (grey areas) is $a = 12.8$ nm according to the FWHM of the intensity distribution, and the distance between the centers of the overlapping cylindrical holes (grey and blue areas) is $d_0 = 8.9$ nm corresponding to the computed splitting (Figure 4c).

First the transmittance and absorptance of hexagonal hole-arrays (Figure 5c, e), which can be produced by homogeneous light illumination (Figure 3a, 5a) were compared to spectra of rectangular hole-arrays (Figure 5d, f), which can be fabricated by IICL in geometry-II (Figure 3b, 5b). In all cases, the presented data are corrected by subtracting the signal of a continuous gold film with analogous thickness. In addition to this, the transmittance spectra were normalized according to the hole area–unit cell area ratio.

The effect of arrays containing stand-alone cylindrical holes, which can be fabricated in gold films via unpolarized light illumination, was compared to the effect of arrays containing hole-doublets, which can be produced by applying polarized light illumination (Figure 5c-f, open-to-closed symbols).

The transmittance and absorptance were computed for all of these hole-arrays first in asymmetric environment, when air surrounding medium is above and inside the holes. Then same arrays were studied in symmetric medium, when NBK7 surrounds the patterned Au film and fills the holes too (Figure 5c, d: green-to-blue; Figure 5e, f: orange-to-red).

Spectral studies were performed for p-polarized light illumination of all hole-arrays in a configuration, where the **E**-field oscillation is perpendicular to the long axis of hole-doublets ($\gamma = 90°$ azimuthal orientation). In addition to this, hole-doublets in symmetric environment were analyzed for the case of **E**-field oscillation parallel to their long axis ($\gamma = 0°$ azimuthal orientation) (Figure 5c, d: closed navy and e, f: closed wine symbols).

Due to the small diameter of individual holes small transmittance and absorptance peaks appear in the neighborhood of gold turnaround frequency in all investigated cases (Figure 5c-f). All maxima observed in absorptance spectra are linked to well-defined transmittance maxima indicating that resonant modes are excited in these spectral intervals (Figure 5c-to-e, and d-to-f) [25].

The extrema depend on the array-geometry and on the surrounding media, too [26]. All extrema are slightly red-shifted in rectangular arrays with respect to the extrema on hexagonal arrays consisting of analogous features in same environment (Figure 5c, e - to - d, f), in accordance with the sensitivity of hole-array's resonance to the periodicity [22].

At the same time, the computation results reveal to strong influence of individual nano-objects shape on the spectral response of arrays (Figure 5c-f). The transmittance and absorptance characteristics indicate that doublets of nano-features, originating from the inherent splitting in intensity distribution in IICL, exhibit unique spectral properties. Namely, the hole-doublets result in a forward shift and in an additional transmittance and absorptance enhancement, when the **E**-field oscillation is perpendicular to their long axes, i.e. under $\gamma = 90°$ azimuthal orientation (Figure 5, closed-to-open symbols). Similar **E**-field orientation preference was observed on spectra of elliptical holes in the literature [27, 28].

The switching from asymmetric to symmetric environment promotes the absorptance and field transportation via coupled modes in both hole-arrays, and results in forward shifted maxima due to the index matching at the opposite metal-dielectric interfaces according to the literature [26, 29, 30].

The corrected-normalized transmittance values are larger than unity in both geometries, only when arrays of hole-doublets are illuminated by light with **E**-field oscillation perpendicular to their long axis, in symmetric environment (Figure 5c, d). Interestingly, minima (534 nm and 540 nm) appear on the transmittance before pronounced maxima (600 nm and 618 nm) in both arrays. Comparison of the schematic drawings shows that both arrays possess the $t$ periodicity, which is commensurate with half-wavelength of the photonic modes at the minima, appearing at spectral positions coincident within the resolution of computations. The coincidence of minima reveals to analogous Bragg-diffraction on hexagonal and rectangular arrays of hole-doublets elongated perpendicularly to the **E**-field oscillation.

The peculiarity of the IICL generated array is that the plasmon wavelength equals with the $p = 300$ nm periodicity in the rectangular unit cell at the transmittance minimum. The fulfillment of the Rayleigh condition might explain the difference between minimal transmissions values observable in the investigated two geometries, which manifests itself in more strongly suppressed transmittance in case of rectangular array.

The pronounced transmittance maxima are observable following these minima (600 nm and 618 nm), i.e. in both complex structures in spectral intervals, where the plasmon wavelengths are larger than the characteristic periodicities. The forward shift of the transmittance maxima with respect to the spectral position, where the Rayleigh condition is fulfilled for propagating plasmonic modes is explained by phase-changes accompanying launching on hole-arrays in the literature [29, 30].

When the **E**-field oscillation is parallel to the long axis of doublets ($\gamma = 0°$), no minima are observable, on the contrary only maxima appear. Comparing the transmittance values in case of two different polarization directions at the minima observed under $\gamma = 90°$, i.e. at positions indicated by symbols with embedded numbers 1 and 3 in Figure 5c, d, we can conclude that larger polarization sensitivity is achievable in case of rectangular array.

The minima appearing on the corrected transmittance of the hole-doublets in symmetric environment at $\gamma = 90°$ indicates that the transmittance on both arrays approximates the transmittance of the continuous film. The near-field investigation revealed considerable **E**-field intensity on the doublets also at these minima (Figure 5a, b, first images).

However, the near-field intensity is the highest on the hole-doublets at the maxima, in accordance with the observed enhanced transmittance and absorptance (Figure 5a, b, second images). This is due to the reinforcement of the external field via induced dipole moments on elongated doublets, when the **E**-field oscillation orientation is perpendicular to their long axis.

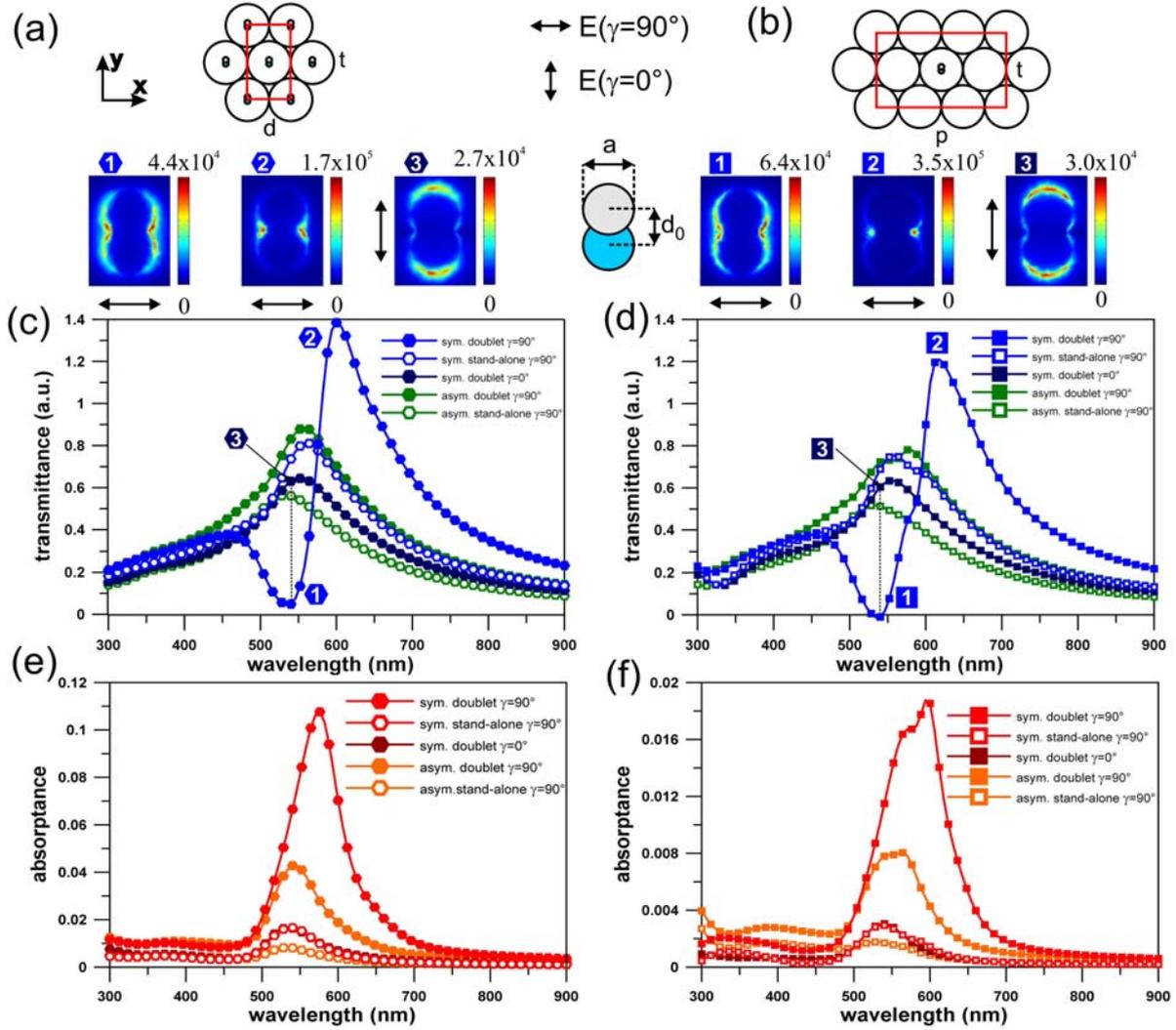

Figure 5. Schematic drawing of unit cells of (a) hexagonal and (b) rectangular arrays studied, and images of the normalized ***E***-field on hole-doublets in case of symmetric media, under $\gamma = 90°$ at the transmittance minima (534 nm and 540 nm) and maxima (600 nm and 618 nm), and under $\gamma = 0°$ orientation at 534 nm and 540 nm wavelengths. (c, d) Corrected-normalized transmittance and (e, f) corrected absorptance spectra of gold films with 45 nm thickness, calculated in presence of different (c, e) hexagonal hole-arrays according to HI in Figure 3a, and (d, f) rectangular hole-arrays according to IICL-II geometry in Figure 3b. The effect of illumination by p-polarized light with incidence plane parallel to x axis ($\gamma = 90°$) is presented for stand-alone holes (open symbols) and hole-doublets (closed symbols) in symmetric (blue, red) and anti-symmetric (green, orange) environments. The doublets in symmetric environment illuminated under $\gamma = 0°$ orientation are presented too (navy, wine closed symbols).

The near-field intensity is considerably smaller in case of $\gamma = 0°$ at the spectral positions corresponding to minima observed on the transmittance under $\gamma = 90°$ azimuthal orientation (Figure 5a, b, last images). This is caused by the smaller local field enhancement available on elongated doublets, when the ***E***-field oscillates along their longer axis [27].

It is a general characteristic of rectangular and hexagonal arrays that in order to maximize the transmission, it is advantageous to apply hole-doublets instead of stand-alone holes and to alter the surrounding media from asymmetric to symmetric. Moreover, these observations prove that IICL-designed rectangular hole-arrays consisting of tiny nano-hole doublets makes possible spectral engineering with enhanced polarization sensitivity due to the involvement of plasmonic phenomena into the EM-field transportation.

## 5. Conclusions

In conclusion, we have shown that IICL, as the integration of interference and colloid sphere lithography, makes possible the near-field concentration of periodic intensity distributions. We have demonstrated that the number of geometrical parameters independently variable with IICL is three or four in two-dimension, depending on the polarization-related sub-structures' dominance. According to our knowledge, IICL is the first parallel method, which makes this high degree of freedom available in complex nano-patterns' fabrication.

We have demonstrated that complex patterns of nano-features might be produced via IICL applying polarized light. Our present study proves that the presence of these complex structures has a pronounced effect on the absorption and transmission spectra of the treated materials, already in case of nano-scaled incorporated features. Moreover, we have shown that spectral engineering is possible with tunable polarization sensitivity via these complex plasmonic patterns.

Further studies are in progress to produce plasmonic nano-clusters by colloid sphere based photolithography techniques, following the method of IICL. Selection of appropriate material and experimental parameters in IICL will enable production of artificial 2D or even 3D complex structures with pre-designed spectral characteristics, or to modify the spectral properties of common materials. Novel types of plasmonic and meta-materials might be fabricated, and very high density data-storage will be possible based on IICL, while the tunable coupled resonances on arrays of nanometric metallic nano-objects will provide more efficient tools for bio-molecular analyses.

**Acknowledgement**

The study was funded by the National Development Agency of Hungary with financial support form the Research and Technology Innovation Funds (CNK-78549, OTKA-NKTH K 75149).

The publication is supported by the European Union and co-funded by the European Social Fund. Project title: "Broadening the knowledge base and supporting the long term professional sustainability of the Research University Centre of Excellence at the University of Szeged by ensuring the rising generation of excellent scientists." Project number: TÁMOP-4.2.2/B-10/1-2010-0012.